\documentstyle[preprint,aps,epsf,floats]{revtex}

\tighten

\def\OMIT#1{{}}
\def\nc{{N_c}}

\def\nc{{N_c}}

\def\yo2{{f_\pi^2}}
\def\llra{{\longrightarrow}}
\def\mapright#1{{\smash{\mathop{\llra}\limits_{#1}}}}

\begin{document}

\preprint{\vbox{
\hbox{INT-PUB-02-51}\hbox{NT@UW-02-035}
}}

\vskip 2.0cm

\title{A Conjecture about Hadrons}
\author{{\bf Silas R.~Beane}}
\address{Institute for Nuclear Theory,
University of Washington, Seattle, WA 98195-1550}

\author{{\bf Martin J.~Savage}}
\address{Department of Physics,
University of Washington, Seattle, WA 98195-1560}

\maketitle

\begin{abstract}
  We conjecture that in the chiral limit of QCD the spectrum of hadrons is
  comprised of decoupled, reducible chiral multiplets. A simple rule is
  developed which identifies the chiral representations filled out
  by the ground-state hadrons. Our arguments are based on the
  algebraic structure of superconvergence relations derived by Weinberg from
  the high-energy behavior of pion-hadron scattering amplitudes.
\end{abstract}

\bigskip
\vskip 4.0cm

\leftline{December 2002}

\vfill\eject

\section{Introduction}
\label{sec:intro}

Understanding the observed structure and decay patterns of the hadrons
continues to be a Holy Grail for nuclear and particle physicists.  A vast
amount of information about hadrons presently exists, and the ongoing
experimental program to measure hadronic properties continues to flourish with
a wealth of high precision data streaming in from Jefferson Laboratory, and
other facilities.  During the past decade or so important progress has been
made toward understanding hadrons from a fundamental perspective with tools
such as the large-$\nc$ limit of QCD~\cite{largeNearly,largeN,largeNapps} and
an increasingly powerful effort in lattice QCD. In this work we return to ideas
about the relation between the high-energy behavior of scattering amplitudes
and the low-energy properties of hadrons developed long ago
in Ref.~\cite{adlerweis,Harari:1966yq,lipkin,gerstein,Harari:1966jz,Fred,Weinberga,Weinbergb}.
These ideas were cast in a modern framework by
Weinberg~\cite{Weinberga,Weinbergb,Weinbergc}, who has long advocated their
central role in hadronic physics.  By comparing models in this framework with
data we are led to conjecture that in the chiral limit, the spectrum of
hadrons is comprised of decoupled, reducible chiral multiplets, and that axial
matrix elements between the states in each reducible chiral multiplet are
determined by the mass spectrum. We develop a simple rule for constructing the
ground-state chiral multiplets of the hadrons.

The observations and arguments we use are conceptually straightforward but
have deep implications. One considers forward pion-hadron scattering in the chiral
limit of QCD, where pions are derivatively coupled to hadrons involved
in any given process via the axial-current operator~\cite{Weinberga}.  The general analytic structure
of the scattering amplitude enables the amplitude at very low energies,
described by an appropriate effective field theory of hadrons, to be related to
the behavior of the scattering amplitude at asymptotically high energies.  
By working in a collinear frame in which the pion-hadron
interaction conserves the helicity of the hadron, and noting
the absence of isospin-one ($I=1$) Regge-trajectories with ${\alpha_1}(0)\ge 1$ in
the crossed $t$-channel~\footnote{
The forward scattering amplitudes have the following asymptotic behavior with $\omega$: 
\begin{eqnarray}
\left({{\cal M}^{(-)\;\lambda}_{\beta h^\prime,\alpha h}}(\omega )\right)_{I=1}&
\mapright{\ \omega\rightarrow\infty}& \ \omega^{\alpha_1 (0)-1}
\ \ ,\ \ 
\left({{\cal M}^{(+)\;\lambda}_{\beta h^\prime,\alpha h}}(\omega )\right)_{I=0,2}
\ \mapright{\ \omega\rightarrow\infty}\ 
\ \omega^{\alpha_{0,2} (0)}
\ \ ,
\nonumber
\label{eq:reggec}
\end{eqnarray}
where $h,h^\prime$ are the isospin indices of the initial and final state
hadron while  $\alpha,\beta$ 
are the isospin indices of the initial and final state pion.
},
one can show that the matrix elements of the axial
current between hadron states of helicity $\lambda$, $X^\alpha_\lambda$, along
with the isospin matrices, $T^\alpha$, form an $SU(2)_L\otimes SU(2)_R$ chiral
algebra,
\begin{eqnarray}
\left[\ T^\alpha\ ,\ X^\beta_\lambda\ \right]
& = & i\epsilon^{\alpha\beta\gamma}\  X^\gamma_\lambda
\ \ ,\ \ 
\left[\ X^\alpha_\lambda \ ,\ X^\beta_\lambda\ \right]
\ =\  i\epsilon^{\alpha\beta\gamma} \ T^\gamma
\ \ \ ,
\label{eq:alg}
\end{eqnarray}
for two-flavor QCD, and $SU(N_f)_L\otimes SU(N_f)_R$ for QCD with $N_f$
flavors~\cite{Weinbergd}. 
In arriving at eq.~(\ref{eq:alg})
it has been assumed that the low-energy amplitude is saturated by
single-particle pole diagrams. The absence of $I=2$ Regge-trajectories with
$\alpha_2(0)\geq 0$ in the crossed $t$-channel allows one to derive algebraic 
relations
involving the axial-current matrix elements and the hadronic mass matrix, $M$,
\begin{eqnarray}
\left[\ X^\alpha_\lambda\ ,\ \left[\ X^\beta_\lambda\ ,\ M^2\ \right]\ 
\right]
& \propto & \delta^{\alpha\beta}
\ \ ,\ \ 
\left[\ X^\alpha_\lambda\ ,\ \left[\ X^\beta_\lambda\ ,\ M J_{x,y}\ 
\right]\ \right]
\ \propto \ \delta^{\alpha\beta}
\ \ \ ,
\label{eq:comma}
\end{eqnarray}
where $J_{x,y}$ are angular-momentum generators in the transverse direction
(where we have chosen $z$ to be the collinear direction). One can show~\cite{Weinberga}
that the first commutator in eq.~(\ref{eq:comma}) implies that $M^2$ is a sum
of two components: one 
transforms as a chiral singlet, $M_{\bf 1}^2$, and one as the isosinglet component
of a $({\bf 2},{\bf 2})$ representation, 
$M_{\bf 22}^2$, of $SU(2)_L\otimes SU(2)_R$. 
The second commutator in eq.~(\ref{eq:comma}) implies that
the same is true for $M J_{x,y}$.  
There is a  further constraint  on the structure of the 
mass matrix arising from the 
empirical fact that the cross section for inelastic diffractive scattering is
significantly smaller than that for elastic scattering.  If the inelastic
scattering cross section were to vanish asymptotically then~\cite{Weinberga}
\begin{eqnarray}
&&
\left[\ M_{\bf 1}^2\ ,\ M_{\bf 22}^2\ \right]
\ =\ 0
\ \ \ .
\label{eq:diff}
\end{eqnarray}
We will present results both with and without this constraint.
{}From the constraints in eq.~(\ref{eq:alg}) and eq.~(\ref{eq:comma})
Weinberg~\cite{Weinbergc} showed that 
\begin{enumerate}
\item[{\bf W1}:] 
any set of hadronic states that furnish a representation of the
  commutation relations in eq.~(\ref{eq:alg}) and eq.~(\ref{eq:comma}) in
  which, for each helicity, any given isospin appears at most once, must be degenerate.
\item[{\bf W2}:] any set of {\it degenerate} hadronic states that furnish a
  representation of the commutation relations in eq.~(\ref{eq:alg}) and
  eq.~(\ref{eq:comma}) also furnish a representation of an $SU(4)\otimes O(3)$
  algebra.
\end{enumerate}

It follows that by assuming only that the $I=J$ tower
of baryon states that naturally arises in the large-$\nc$ limit of QCD 
saturates
the commutation relations in eq.~(\ref{eq:alg}) and eq.~(\ref{eq:comma}), the
states in the tower must be degenerate and the axial matrix elements must be
those of the naive constituent quark model (NCQM)~\footnote{ By NCQM we mean
  the barest form of the constituent quark model (CQM) where there are no
  spin-dependent interaction between quarks arising from either the
  phenomenological ``gluon exchange''~\cite{DeRu} or 
``pseudo-Goldstone boson exchange''~\cite{Mano,Gloz} as
  in the chiral quark model, i.e.  the $\Delta$ and nucleon are degenerate.}.
By construction they exhibit a spin-flavor $SU(4)\otimes O(3)$ symmetry
(which becomes a contracted $SU(4)\otimes O(3)$ symmetry in the
large-$\nc$ limit~\cite{largeN}). 
The ground-state chiral multiplet has been constructed
for the $I=J$ tower and shown to give the (contracted) spin-flavour $SU(4)$ 
results~\cite{Beane4}.

These are indeed beautiful results and allow one to understand how the
NCQM can provide a rigorous mnemonic for describing properties of hadrons in
the chiral limit with a single assumption about the hadronic spectrum.
However, there are certain aspects of this construction that we wish to 
investigate further. 
First, if we want to understand hadrons in QCD with $\nc=3$, we need to 
consider schemes which saturate the commutation relations in eq.~(\ref{eq:alg}) and
eq.~(\ref{eq:comma}) with more than just the ground-state $I=J$ tower.
Put another way, if we want to understand $1/\nc$ corrections to the
picture described above then we must consider more complicated saturation
schemes that allow mixings between the large-$\nc$ tower states (which for
$\nc=3$ contains only the nucleon and $\Delta$) and other
states~\cite{Weinbergc,Beane4}.  Second, there are non-zero mass splittings between the low-lying
baryons that will survive in the chiral limit which must appear in any
consistent description.  
Third, while the ratio of axial matrix elements of the
NCQM agree reasonably well with available experimental data, 
their absolute values are too large by $\sim 30\%$.

\section{The Light Baryons}
\label{sec:lightB}

Consider the lowest-lying baryons with valence structure composed of up 
and down quarks only, such as the nucleon and the $\Delta$-resonance. 
The only representations of $SU(2)_L\otimes SU(2)_R$ that contain only
$I={1\over 2}$ and $I={3\over 2}$ states are
$({\bf 1},{\bf 2})$, $({\bf 2},{\bf 1})$, 
$({\bf 1},{\bf 4})$, $({\bf 4},{\bf 1})$, 
$({\bf 2},{\bf 3})$ and $({\bf 3},{\bf 2})$.
Given one's bias from the NCQM, it is natural to consider chiral multiplets that are completely
filled out by the nucleon and the $\Delta$-resonance alone. 
Clearly there is only a small number of
chiral-multiplets (for each helicity state) that these hadrons can belong to,
without admixtures of other states.
For the $\lambda={1\over 2}$ helicity states they are 
$({\bf 1},{\bf 2})\oplus ({\bf 1},{\bf 4})$, $({\bf 2},{\bf 1})\oplus ({\bf 1},{\bf 4})$, 
$({\bf 1},{\bf 2})\oplus ({\bf 4},{\bf 1})$, 
$({\bf 2},{\bf 1})\oplus ({\bf 4},{\bf 1})$, 
$({\bf 2},{\bf 3})$ and 
$({\bf 3},{\bf 2})$,
while for the $\lambda={3\over 2}$ helicity states they are $({\bf 1},{\bf 4})$
and $({\bf 4},{\bf 1})$.  
Given the results of {\bf W1} and {\bf W2} it is 
clear that the only mass spectrum possible for all of these chiral embeddings,
either reducible or irreducible, is one in which the nucleon and $\Delta$ are
degenerate, and have axial matrix elements that are those of the NCQM.
Clearly the degenerate masses and the excessively-large 
axial-current matrix elements render these 
multiplet structures insufficient. 
However, from the NCQM point of view one uses this as a
starting point and perturbatively includes contributions that appear somewhat
natural, such as spin-dependent quark interactions and the quenching of the
constituent quark axial coupling from $g_A^{(q)}=1$ to some lesser value to reproduce
the nucleon axial matrix elements. In deriving the algebraic
constraints, the analysis has
been performed in the chiral limit and single-particle contributions have
been taken to saturate the commutators.  Given the small value of the pion mass and the
extensive studies in chiral perturbation theory in the nucleon-$\Delta$
sector~\cite{JM,Ulf}, the quark masses will not bring the NCQM chiral multiplet structure
into agreement with data. Furthermore, the saturation of the commutators with
single-particle states is precisely the assumption that one makes in constructing
the low-energy effective field theory description of the $\pi N\Delta$ system, 
and this construction is very successful.
Thus, one is led to conclude that the NCQM
chiral multiplet structure is incomplete and likely
not a sensible basis for a perturbative expansion.

To determine the minimal realistic chiral multiplet structure,
consider the isospin content of the QCD interpolating fields that have non-zero 
overlap with the nucleon or $\Delta$,
\begin{eqnarray}
\varepsilon_{abc}\ q^a\  q^b\  q^c\ \rightarrow \ \textstyle{1\over 2}^+\oplus
\textstyle{1\over 2}^+\oplus \textstyle{3\over 2}^+
\ \ \ ,
\label{eq:interheavies}
\end{eqnarray}
where $a,b,c$ are color indices and all the Dirac and flavour indices have been
suppressed. We have used the schema $I^P$ where $P$ is parity.
This interpolating field contains an additional $I={1\over 2}$
baryon beyond the nucleon and $\Delta$. 

\vskip0.2cm

\noindent{\it We conjecture that a chiral
  multiplet which describes the low-lying hadrons is the minimal chiral
  representation which includes all of the isospin multiplets in the QCD
  interpolator for that hadron at least once, and for which there is no
  degeneracy within the multiplet, unless required by an additional symmetry
  like heavy-quark symmetry or flavor $SU(3)$ symmetry.}  

\vskip0.2cm

Consider the ground-state chiral multiplet for 
the $\lambda={1\over 2}$ baryons. From eq.~(\ref{eq:interheavies}) we require a representation
that contains at least two $I={1\over 2}$ states and one
$I={3\over 2}$ state, all of like-parity. 
In order to avoid degeneracy within the multiplet while allowing
pion transitions we require a representation with nonvanishing matrix elements of
$M_{\bf 22}^2$. Hence the ground-state chiral
multiplet for the $\lambda={1\over 2}$ baryons is uniquely determined by the
conjecture to be $({\bf 2},{\bf 3})\oplus ({\bf 1},{\bf 2})$~\footnote{
  Parity interchanges $SU(2)_L$ and $SU(2)_R$ representations. Therefore
  if we assign the 
  $\lambda=+{1\over 2}$ states to an $({\bf 2},{\bf 3})\oplus({\bf 1},{\bf 2})$
  representation, parity requires that the
  $\lambda=-{1\over 2}$ states are in the  $({\bf 3},{\bf 2})\oplus({\bf 2},{\bf 1})$
  representation~\cite{Weinberga}.}. This multiplet
was considered by Weinberg~\cite{Weinberga} (and also by Gilman and
Harari~\cite{Fred}).  As discussed in detail below, the additional baryon in
the chiral multiplet is identified as the Roper resonance, $N(1440)$~\cite{Ubi}.  The
actual Dirac and lorentz structure of the interpolating fields is not at all
obvious to us. An early discussion of this problem can be found in work by
Casher and Susskind~\cite{CS74} and a recent discussion can be found in
Ref.~\cite{Beane2}.

The helicity states of the nucleons, $N$,
and excited nucleons, $N^\prime$, are in $I={1\over 2}$ representations of
$SU(2)_I$, described by a tensor with a single fundamental index.  Likewise,
the helicity states of the $\Delta$'s and the excited $\Delta$'s are in 
$I={3\over 2}$ representations of $SU(2)_I$, 
described by a symmetric tensor with three
fundamental indices. 
We will now construct the $({\bf 2},{\bf 3})\oplus ({\bf 1},{\bf 2})$
representation which contains $N$, $N^\prime$ and $\Delta$.
At leading order (LO) in the chiral expansion the axial matrix elements are
defined through the currents~\cite{JM}
\begin{eqnarray}
&& J^{\alpha, 5}_{{\uparrow},LO}
 =  g_A\ N_\uparrow^\dagger \ T^\alpha\  N^{}_\uparrow +
g_A^\prime\ \left(\ N_\uparrow^\dagger \ T^\alpha\  N^\prime_\uparrow 
\ +\ {\rm h.c.}\ \right)
\ +\ g^{\prime\prime}_A\ N_\uparrow^{\prime\dagger} \ T^\alpha\  N^\prime_\uparrow \nonumber \\
&&\ -\ {\cal C}_{\Delta N} \ \left(\ \sqrt{\textstyle{2\over 3}} N_\uparrow^\dagger \ T^\alpha\  
\Delta^{}_\uparrow \ +\ {\rm h.c.}\ \right)
\ -\ {\cal C}_{\Delta N^\prime}\ \left(\ \sqrt{\textstyle{2\over 3}} 
N_\uparrow^{\prime\dagger} \ T^\alpha\ \Delta^{}_\uparrow 
\ +\ {\rm h.c.}\ \right) \nonumber \\
&&\ -\ {\cal H}_{\Delta\Delta} \textstyle{1\over 3} 
\Delta_\uparrow^\dagger\ T^\alpha\
\Delta^{}_\uparrow \ \ , \nonumber \\
&& J^{\alpha, 5}_{{\Uparrow},LO}
 =  
\ -\ {\cal H}_{\Delta\Delta}\ \Delta_\Uparrow^\dagger\ T^\alpha\
\Delta^{}_\Uparrow
\label{eq:chiptaxial}
\end{eqnarray}
for the $\lambda={1\over 2}$ helicity states ($\uparrow$) and
$\lambda={3\over 2}$ helicity states ($\Uparrow$), respectively.
The NCQM places the $N$ and $\Delta$ in the $\bf{20}$-dimensional
representation of spin-flavor $SU(4)$, and
the $N^\prime$ and a $\Delta^\prime$ in the $\bf{20}^\prime$ 
representation. This leads to the familiar NCQM predictions: 
$g_A=g_A^{\prime\prime}={5\over 3}$, $g^\prime_A=0$,
${\cal C}_{\Delta N}=-2$ and ${\cal H}_{\Delta\Delta}=-3$.

In order to construct the ground-state baryon chiral multiplet
consistent with our conjecture,
we introduce the fields $S_a$, $T_{a,bc}$ to include the 
$\lambda=+{1\over 2}$ helicity states and the field 
$D_{abc}$ to include the $\lambda=+{3\over 2}$ helicity states.
The field $S_a$ 
transforms as $({\bf 1},{\bf 2})$ under $SU(2)_L\otimes SU(2)_R$; 
that is, $S\rightarrow LS$,
while the field $D_{abc}$ 
transforms as $({\bf 1},{\bf 4})$ under $SU(2)_L\otimes SU(2)_R$; 
that is,
$D\rightarrow LLL D$.
It is straightforward to embed an $I={1\over 2}$ and an $I={3\over 2}$ 
state into a single
irreducible representation of $SU(2)_L\otimes SU(2)_R$, the 
$({\bf 2},{\bf 3})$.
The field $T_{a,bc}$ transforms as $T\rightarrow RLLT$, and 
in terms of fields transforming as $I={1\over 2}$, $S_T$, 
and $I={3\over 2}$, $D_T$, $T$ can be written as
\begin{eqnarray}
T_{a,bc} & = & 
{1\over\sqrt{6}}\left(\ S_{T,b}\ \epsilon_{ac}\ +\ S_{T,c}\ \epsilon_{ab}\
\right)
\ +\ D_{T,abc} 
\ \ \ .
\end{eqnarray}
We also introduce a spurion field, $v^a_b$, which
transforms as $ v\ \rightarrow\  L\  v\  R^\dagger $, such that
$\langle {v^a_b} \rangle =M_{{\bf 2}{\bf 2} }^2\ \delta^a_b$.
The free-field dynamics of the helicity states
are determined by the two-dimensional effective Lagrange densities
constructed from the available tensors,
\begin{eqnarray}
{\cal L}_{\,\uparrow} & = &
\partial_+ T^{a,bc\dagger} \partial_- T_{a,bc}\ +\ 
\partial_+ S^{a\dagger} \partial_- S^{}_{a}
\ -\ M_{{\bf 1}T}^2\   T^{a,bc\dagger} T_{a,bc}
\ -\ M_{{\bf 1}S}^2\   S^{a\dagger} S^{}_{a}
\nonumber\\
& &\qquad\quad 
-\ {\cal A}\ \left(\  T^{a,bc\dagger} v^{d\dagger}_a S_{b} \epsilon_{cd}
\ +\ {\rm h.c.}\ \right) \ ,
\nonumber\\
{\cal L}_{\,\Uparrow} & = & \partial_+ D^{a,bc\dagger} \partial_- D_{a,bc}
\ -\ M_{{\bf 1}D}^2\   D^{a,bc\dagger} D_{a,bc}
\ ,
\end{eqnarray}
where ${\cal A}$ is an undetermined parameter 
and $x_{\pm} =z\pm t$ with $z$ the collinear direction. 
Notice that the helicity components of the baryons act as scalar
fields. The current operators
that satisfy the constraints imposed by 
eq.~(\ref{eq:alg}) and eq.~(\ref{eq:comma})
take the form
\begin{eqnarray}
\hat T_\uparrow^\alpha 
& = & \ T^{a,bc\dagger} \left( T^\alpha \right)_a^{d} T_{d,bc}
\ + \ 2\  T^{a,bc\dagger} \left( T^\alpha \right)_b^{d} T_{a,dc}
\ +\  S^{a\dagger} \left( T^\alpha \right)_a^{d} S_d \ ,
\nonumber\\ 
\hat X_\uparrow^\alpha 
& = & \ T^{a,bc\dagger} \left( T^\alpha \right)_a^{d} T_{d,bc}
\ - \ 2\  T^{a,bc\dagger} \left( T^\alpha \right)_b^{d} T_{a,dc}
\ -\  S^{a\dagger} \left( T^\alpha \right)_a^{d} S_d \ ,\nonumber\\ 
\hat T_\Uparrow^\alpha 
& = & \ 3\ D^{abc\dagger} \left( T^\alpha \right)_a^{d} D_{dbc}\ ,
\nonumber\\ 
\hat X_\Uparrow^\alpha 
& = & \ -3\ D^{abc\dagger} \left( T^\alpha \right)_a^{d} D_{dbc}
\ \ .
\end{eqnarray}
The mass eigenstates are linear combinations of the chiral
eigenstates with a mixing angle $\psi$. Setting
$M^2_{{\bf 1}T}=M^2_{{\bf 1}D}$ one can easily check that the commutators of
eq.~(\ref{eq:alg}) and eq.~(\ref{eq:comma}) are satisfied.
Diagonalizing the mass matrix and matching to the 
chiral perturbation theory current in eq.~(\ref{eq:chiptaxial})
leads to
\begin{eqnarray}
& &g_A\ =\ 1\ +\ \textstyle{2\over 3}\cos^2\psi
\ \ ,\ \ 
g^\prime_A\ = \textstyle{2\over 3}\sin\psi\cos\psi
\ \ ,\ \ 
g^{\prime\prime}_A\ =\ 1\ +\ \textstyle{2\over 3}\sin^2\psi\ , \nonumber\\ 
& &\qquad{\cal C}_{\Delta N}\ =\ -2 \cos\psi
\ \ ,\ \ 
{\cal C}_{\Delta N^\prime}\ =\ -2 \sin\psi
\ \ ,\ \ 
{\cal H}_{\Delta\Delta}\ =\ -3 \ , \nonumber\\ 
& &\qquad\qquad  M_N^2\cos^2\psi
\ +\ M_{N^\prime}^2\sin^2\psi\  =\  M_\Delta^2 
\ ,
\label{eq:axials}
\end{eqnarray}
where $\psi$ is the mixing angle between the two $I={1\over 2}$ 
multiplets~\footnote{We believe the mass relation in Ref.~\cite{Weinberga}
to be incorrect.}.
If we further impose the inelastic diffraction constraint, we find that
$M_{{\bf 1}T}^2 = M_{{\bf 1}S}^2$ and 
consequently $\psi={\pi\over  4}$, which
corresponds to maximal mixing~\footnote{
This choice of the mixing angle corresponds to a discrete
symmetry of the free lagrange density which interchanges $S$ and $S_T$~\cite{Beane2}. In all
cases we study in this paper, the constraint of no inelastic diffraction
corresponds to a discrete symmetry of the collinear field theory.}.
This then gives
\begin{eqnarray}
& &\quad\qquad g_A\ =\ \textstyle{4\over 3}
\ \ \ ,\ \ \
g^\prime_A\ = \textstyle{1\over 3}
\ \ \ ,\ \ \
g^{\prime\prime}_A\ =\  \textstyle{4\over 3}  \ , \nonumber\\ 
& &{\cal C}_{\Delta N}\ =\ -\sqrt{2} 
\ \ ,\ \ 
{\cal C}_{\Delta N^\prime}\ =\ -\sqrt{2} 
\ \ ,\ \ 
{\cal H}_{\Delta\Delta}\ =\ -3 \ , \nonumber\\ 
& &\quad\qquad\qquad  M_\Delta^2\ -\ M_N^2\  \ =\ M_{N^\prime}^2\ -\ M_\Delta^2 \ .
\label{eq:axialswithdiff}
\end{eqnarray}
These values are impressively close to those in nature and it
is conceivable that the agreement may improve as the physical values are
extrapolated to the chiral limit. Using the nucleon and $\Delta$ masses as
input one finds $M_{N^\prime}=1467~{\rm MeV}$, consistent with the Roper
resonance. Notice that both $g_A$ and ${\cal C}_{\Delta N}$ are decreased from their NCQM
in the direction of experiment (see Table~\ref{tab2}).
The phenomenology of the axial couplings in this scenario 
is discussed in detail in Ref.~\cite{Ubi}.
Recent work on $\pi N\rightarrow\pi\pi N$ scattering by Fettes~\cite{Nadia} 
has determined a range for the higher-order contributions to the axial current 
in chiral perturbation theory, described by the constant $\overline{d}_{16}$.
The range of values for $\overline{d}_{16}$ is consistent with a value 
of $g_A={4\over 3}$ in the chiral limit.
Finally, we point out that it is likely that
since the Roper-nucleon mass splitting is less than the chiral symmetry
breaking scale, the non-vanishing quark-mass corrections to this chiral multiplet can
be computed using chiral perturbation theory~\cite{SavageBeane}.

\begin{table}[htb] 
\begin{tabular}{||ccccc||}
   &  CONJECTURE & CONJECTURE* & NCQM  & EXPERIMENT  \\
\hline  \hline
\rule[-2mm]{0mm}{6mm} $|g_A|$  & $1 + \textstyle{2\over 3}\cos^2\psi$ &
${4\over 3}$  
& ${5\over 3}$ & $1.26$  \\
    \hline
\rule[-2mm]{0mm}{6mm} $|{\cal C}_{\Delta N}|$  & $2\cos\psi$ & $\sqrt{2}$ & $2$ & $1.2\pm0.1$ \\
    \hline
\rule[-2mm]{0mm}{6mm} $|{\cal H}_{\Delta\Delta}|$ & $3$ & $3$ & $3$ & $2.2\pm 0.6$  \\
    \hline
    \hline
\rule[-2mm]{0mm}{6mm} $|g_2|$  & $2$ & $2$ & $2$ & $--$  \\
    \hline
\rule[-2mm]{0mm}{6mm} $|g_3|$  & $\sqrt{2}\cos\psi $ & $1$ & $\sqrt{2}$ & $0.95\pm 0.08\pm 0.08$  \\
    \hline
    \hline
\rule[-2mm]{0mm}{6mm} $|g|$   &  $\cos (\theta +\phi )$ & $0$ & $1$ & $0.27\pm 0.03 \pm 0.04$  \\
    \hline
\rule[-2mm]{0mm}{6mm} $|h|$  & $\sin (\theta +\phi )$ & $1$ & $0$ & $--$  
\end{tabular}
\caption{Axial couplings for the light baryons, heavy baryons
  and heavy mesons. 
The third and second columns give the predictions of
the chiral conjecture both with (*) and without the
  inelastic diffraction constraint of eq.~(\ref{eq:diff}).
The fourth column gives the axial couplings of the NCQM.
The experimental values have been determined via branching fractions that
appear in the particle data group~\protect\cite{pdg02}.
The extractions of ${\cal C}_{\Delta N}$ and
${\cal H}_{\Delta\Delta}$ from data were made in $SU(3)$ chiral perturbation theory~\protect\cite{BSS}.
For a discussion of the experimental value of $g$ and other allowed values, see the text.}

\label{tab2}
\end{table}

\section{The Heavy Baryons}
\label{sec:heavyB}

For baryons containing one heavy quark one naturally considers the NCQM states as
being partners, such as the $\Lambda_c^+$, 
the $\Sigma_c^{++,+,0}$ and the $\Sigma_c^{* ++,+,0}$
in the charmed sector.
The minimal chiral multiplet structure for these states gives a continuous
spectrum of axial-current matrix elements when the algebraic constraints 
of eq.~(\ref{eq:alg}) and eq.~(\ref{eq:comma})
are
imposed.  However, also requiring heavy-quark symmetry (HQS)
yields the axial coupling constants of the NCQM.
An interpolating field for the heavy baryons takes the form
\begin{eqnarray}
\varepsilon_{abc}\ Q^a\  q^b\  q^c \ \rightarrow \ 0^+\oplus 1^+ \ \ \ ,
\label{eq:inter}
\end{eqnarray}
where $Q$ denotes a heavy quark~\footnote{The discussion also holds for strange
baryons such as the $\Lambda$ or $\Sigma$.}
and again we have used the schema $I^P$ where $P$ is parity.
In order to describe the $\Lambda_Q$, the $\Sigma_Q$ and the $\Sigma_Q^*$
two copies of the heavy-baryon interpolator must be present, and thus a
$\Lambda_Q^\prime$ will be present in the ground-state chiral multiplet. 
The unique chiral multiplet consistent with our conjecture is
$({\bf 2},{\bf 2})\oplus ({\bf 1},{\bf 3})\oplus ({\bf 1},{\bf 1})$.
One additional ingredient in the heavy-quark sector that is absent in the 
light-quark sector is heavy quark spin-symmetry (HQSS).
In the experimentally-determined heavy-baryon spectrum, the 
$\Sigma_Q$ and $\Sigma_Q^*$ are identified as members of an irreducible
representation of HQSS, and become degenerate in the heavy quark limit.
HQSS greatly simplifies the form of the axial-current matrix elements
and one finds~\cite{Cho}
\begin{eqnarray}
J^{\alpha,5}_{Q\uparrow,LO} & = & 
g_2\ \left(\ 
\textstyle{2\over 3} {\rm Tr}\left[\Sigma_{Q\uparrow}^\dagger T^\alpha \Sigma_{Q\uparrow}
\right]
\ +\ 
\textstyle{1\over 3} {\rm Tr}\left[\Sigma_{Q\uparrow}^{*\dagger} T^\alpha 
\Sigma_{Q\uparrow}^*\right]
\ +\ 
\textstyle{\sqrt{2}\over 3} \left(\ 
{\rm Tr}\left[\Sigma_{Q\uparrow}^{*\dagger} T^\alpha \Sigma_{Q\uparrow}\right]
\ +\ {\rm h.c.}\ \right)\ \right)
\nonumber\\
& + & 
g_3\ \left(\ 
\textstyle{1\over\sqrt{3}}\ \Lambda_{Q\uparrow}^\dagger 
{\rm Tr}\left[ T^\alpha \Sigma_{Q\uparrow}\right]
\ -\ 
\textstyle\sqrt{2\over 3}\ \Lambda_{Q\uparrow}^\dagger 
{\rm Tr}\left[ T^\alpha \Sigma_{Q\uparrow}^*\right]
\ +\ {\rm h.c.}
\ \right)
\nonumber\\
& + & 
g_3^\prime\ \left(\ 
\textstyle{1\over\sqrt{3}}\ \Lambda_{Q\uparrow}^{\prime\dagger} 
{\rm Tr}\left[ T^\alpha \Sigma_{Q\uparrow}\right]
\ -\ 
\textstyle\sqrt{2\over 3}\ \Lambda_{Q\uparrow}^{\prime\dagger} 
{\rm Tr}\left[ T^\alpha \Sigma_{Q\uparrow}^*\right]
\ +\ {\rm h.c.}
\ \right) \ \ ,
\nonumber\\
J^{\alpha,5}_{Q\Uparrow,LO} & = & 
g_2\ 
{\rm Tr}\left[\Sigma_{Q\Uparrow}^{*\dagger} T^\alpha 
\Sigma_{Q\Uparrow}^*\right]
\ \ \ .
\label{eq:HQaxialshel}
\end{eqnarray}
The NCQM values for the axial couplings are 
$g_2= 2$ and $g_3= -\sqrt{2}$~\cite{Cho,YCCLLYa}.

We will now construct the 
$({\bf 2},{\bf 2})\oplus ({\bf 1},{\bf 3})\oplus({\bf 1},{\bf 1})$
representation that contains $\Lambda_Q$, $\Lambda_Q^\prime$,
$\Sigma_Q$ and the $\Sigma_Q^*$.
We introduce the fields
$\lambda_1$, $Z$ and $Y$ to describe the $\lambda={1\over 2}$ helicity states
and the field $Q$ to describe the $\lambda={3\over 2}$ helicity state.
$\lambda_1$ transforms as $({\bf 1},{\bf 1})$ under the chiral group,
$Z^a_b$ transforms as $({\bf 1},{\bf 3})$ under the chiral group,
$Z\rightarrow L Z L^\dagger$, and  
$Y^a_b$ transforms as $({\bf 2},{\bf 2})$ under the chiral group,
$Y\rightarrow L Y R^\dagger$, with 
$Y_0$ the $I=0$ component, and $Y_1$ the 
$I=1$ component.
The field $Q^a_b$ transforms as $({\bf 1},{\bf 3})$ under the chiral group,
$Q\rightarrow L Q L^\dagger$.
The free-field dynamics of the helicity states
are determined by the two-dimensional effective Lagrange densities
constructed from $\lambda_1$, $Z$ and $Y$,
\begin{eqnarray}
{\cal L}_\uparrow & = & 
\partial_+\lambda_1^\dagger \partial_-\lambda_1
\ +\ 
{\rm Tr}\left[\ \partial_+ Y^\dagger \partial_- Y\ \right]
\ +\ 
{\rm Tr}\left[\ \partial_+ Z^\dagger \partial_- Z\ \right]
- M_{{\bf 1}Y}^2 \ {\rm Tr}\left[\ Y^\dagger Y\ \right]
\nonumber\\
& - &  M_{{\bf 1}Z}^2 \ {\rm Tr}\left[\ Z^\dagger Z\ \right]
\ -\ M_{{\bf 1}\lambda_1}^2 \ \lambda_1^\dagger\lambda_1
\ -\ \left(\ 
{\cal A}_1\ {\rm Tr}\left[\ Y^\dagger Z\ v\ \right]
\ +\ {\cal A}_2\ {\rm Tr}\left[\ Y^\dagger\ v\ \right]\ \lambda_1
\ +\ {\rm h.c.}\ \right) \ \ ,
\nonumber\\
{\cal L}_\Uparrow & = & 
{\rm Tr}\left[\ \partial_+ Q^\dagger \partial_- Q\ \right]
- M_{{\bf 1}Q}^2 \ {\rm Tr}\left[\ Q^\dagger Q\ \right]
\ \ \ ,
\end{eqnarray}
where ${\cal A}_{1,2}$ are unknown constants.
The $\Lambda_Q^{(\prime)}$ mass eigenstates are linear combinations of the 
$\lambda_1$ and $Y_0$ fields with a mixing angle $\psi$, while the 
$\Sigma_Q^{(*)}$ mass eigenstates are linear combinations of the 
$Z$ and $Y_1$ fields, with mixing angle $\phi$.
The constraints imposed by 
eq.~(\ref{eq:alg}) and eq.~(\ref{eq:comma}) along with HQSS 
have a non-trivial solution for the axial coupling constants
and masses
\begin{eqnarray}
&& g_2\ =\ 2
\ \ \ ,\ \ \ 
g_3\ =\ -\sqrt{2}\cos\psi
\ \ \ ,\ \ \ 
g_3^\prime\ =\ -\sqrt{2}\sin\psi \ ,\nonumber \\
&&\quad\qquad
M_{\Lambda_Q}^2\cos^2\psi\ +\ 
M_{\Lambda_Q^\prime}^2\sin^2\psi\ =\ 
M_{\Sigma_Q^{(*)}}^2
\ \ \ .
\end{eqnarray}
Imposing the constraint in eq.~(\ref{eq:diff}) arising from the 
absence of inelastic diffractive scattering requires $\psi={\pi\over 4}$ 
which gives
\begin{eqnarray}
&&\quad g_2\ =\ 2
\ \ \ ,\ \ \ 
g_3\ =\ g_3^\prime\ =\ -1 \ ,\nonumber \\
&&
M_{\Sigma_Q^{(*)}}^2\ -\ M_{\Lambda_Q}^2 \ =\ 
M_{\Lambda_Q^\prime}^2 \ -\ M_{\Sigma_Q^{(*)}}^2
\ \ \ ,
\end{eqnarray}
or $\psi=0$, which decouples the $\Lambda_c^\prime$ as a stand-alone
$({\bf 1},{\bf 1})$ representation and recovers the NCQM values for
the heavy-baryon multiplet axial transitions ($g_2= 2$, $g_3= -\sqrt{2}$ and
$g_3^\prime=0$). 

Information continues to be accumulated about the properties and decays
of baryons containing heavy quarks
(for a recent review of experimental data see Ref.~\cite{yelton02,Artuso:2000xy},
and references therein and also the particle data group~\cite{pdg02}).
Ideally, one would like to have detailed information on b-baryons so as to be as
close to the HQ limit as nature will allow.
However, there is much more information
on charmed baryons, simply due to the number that have been produced 
in the laboratory.
Experimentally, the measured width of the $\Sigma_c^{*++}$ is
$\Gamma (\Sigma_c^{*++}) = 17.9^{+3.8}_{-3.2}\pm 4.0~{\rm MeV}$, which
fixes the axial coupling $g_3$ to be $|g_3|=0.95\pm 0.08\pm 0.08$.
This value is significantly smaller than the NCQM value of $|g_3|=\sqrt{2}$,
but consistent with $|g_3|=1$  that one obtains with $\psi={\pi\over 4}$. 
Thus our conjecture predicts there to be a chiral partner to the 
$\Lambda_c$, $\Sigma_c$ and the $\Sigma_c^*$, the positive-parity $\Lambda_c^\prime$
with a mass of $M_{\Lambda_c^\prime}\sim 2688~{\rm MeV}$ that is the analogue
of the Roper in the light baryon sector and is not yet observed.
The axial coupling between the $\Lambda_c^\prime$ and the 
$\Sigma_c$ and the $\Sigma_c^*$ is predicted to be 
$|g_3^\prime|=1$. The fact that the charmed spectrum is not as close to the
heavy-quark limit as one would like means that the $\Lambda_c^\prime$ mass
may be somewhat heavier than we have estimated here.

\section{The Light Mesons}
\label{sec:lightM}

In the case of the light mesons,
our conjecture has been proven to be true in the large-$\nc$ limit of
QCD~\cite{Weinbergd} 
using eq.~(\ref{eq:alg}), eq.~(\ref{eq:comma}) and 
eq.~(\ref{eq:diff}). The $\lambda=0$
helicity states of the $\pi$, $\rho$, $f_0$ and $a_1$ form a chiral quartet
that is decoupled from the other mesons. This provides a QCD-based
interpretation of the work by Gilman and Harari~\cite{Fred} in which it was
shown that the complete set of Adler-Weisberger sum-rules in the pion sector could be satisfied
with just these four meson states.  The one free mixing angle
introduced by Gilman and Harari is
predicted to be $\psi={\pi\over 4}$ and gives a $\rho$ width
that is consistent with experiment. We now demonstrate how
these results are implied by our conjecture. In effect, we will see that our conjecture recovers
precisely the full saturation scheme of Gilman and Harari~\cite{Fred}.
The interpolating fields for the mesons are of the form
$\overline{q}^{\,a} q_a$ and contain isospin representations $I=0\oplus 1$ only.
The chiral multiplets which can give rise to these isospins and no others are
$({\bf 1},{\bf 1})$, $({\bf 1},{\bf 3})$, $({\bf 3},{\bf 1})$,
and $({\bf 2},{\bf 2})$. 
As the  mesons have $\lambda=0$ helicity states
{\it normality}, $\eta\equiv P(-1)^J$, constrains the structure of the
mass matrix and axial currents. Moreover, $G$-parity adds a new
complication as there is an additional symmetry, $G\eta$, which commutes with
the axial operator, $\hat X_\alpha$. Hence one should consider sectors with
$G\eta=+1$ and $G\eta=-1$ {\it separately}.
Meson interpolators with space-time quantum numbers, $J^{PC}$, of
pseudoscalar, vector, axialvector and scalar character decompose
to
\begin{eqnarray}
\left(\ \overline{q}^{\, a} q_a\ \right)_{P}\ &\rightarrow &\ 0^-_-\ \oplus\ 1^-_+ \ \ , \ \ 
\left(\ \overline{q}^{\, a} q_a\ \right)_{V}\ \rightarrow\ 0^+_-\ \oplus\ 1^+_+
\ , \nonumber \\
\left(\ \overline{q}^{\, a} q_a\ \right)_{A}\ & \rightarrow &\ 0^-_-\ \oplus\ 1^-_+\ \ , \ \ 
\left(\ \overline{q}^{\, a} q_a\ \right)_{S}\ \rightarrow\ 0^+_+\ \oplus\ 1^+_- \ ,
\label{eq:lightmesonsINTS}
\end{eqnarray}
where we use the schema $I^\eta_{G\eta}$.
Consider first the sector with $G\eta=+1$ which includes the pion. 
Since normality interchanges
$({\bf 1},{\bf 3})$ and $({\bf 3},{\bf 1})$,
the minimal reducible representation that allows for non-zero mass splittings
is $({\bf 2},{\bf 2})\oplus ({\bf 1},{\bf 3})\oplus ({\bf 3},{\bf 1})$
which corresponds to $1^-_+$, $1^-_+$, $1^+_+$ and $0^+_+$ from eq.~(\ref{eq:lightmesonsINTS}).
(This representation is denoted by $v\oplus t$ in Ref.~\cite{Weinberga}.)
Hence the pion belongs to a chiral representation composed of the $\lambda=0$
helicity states of $\pi$, $f_0(600)$, $\rho(770)$ and $a_1(1260)$.
We introduce the fields $h$, $k$ and $t$ which transform as
$h\rightarrow LhL^\dagger$,
$k\rightarrow RkR^\dagger$ and 
$t\rightarrow LtR^\dagger$ under chiral transformations, and 
whose free-field dynamics are described by the Lagrange density
\begin{eqnarray}
{\cal L}_0^+ & = & 
{\rm Tr}\left[\ \partial_+ t^\dagger \partial_- t\ \right]
\ +\ 
{\rm Tr}\left[\ \partial_+ h^\dagger \partial_- h\ \right]
\ +\ 
{\rm Tr}\left[\ \partial_+ k^\dagger \partial_- k\ \right]
- M_{{\bf 1}t}^2 \ {\rm Tr}\left[\ t^\dagger t\ \right]
\nonumber\\
&&
- M_{{\bf 1}+}^2 \ {\rm Tr}\left[\ (h+k)^\dagger (h+k)\ \right]
- M_{{\bf 1}-}^2 \ {\rm Tr}\left[\ (h-k)^\dagger (h-k)\ \right]
\nonumber\\
&&
\ -\ \left(\ {\cal A}\ {\rm Tr}\left[\ t^\dagger ( h v\ +\ v^\dagger k )\ \right]
\ +\ {\rm h.c.}\ \right)
\ \ \ ,
\end{eqnarray}
where ${\cal A}$ is an unknown constant and the current operators are
\begin{eqnarray}
&&
{\hat T}_0^\alpha\ =\ {\rm Tr}\left[\ t^\dagger T^\alpha t\ -\ t^\dagger t T^\alpha\ \right]
\ +\ 
2\ {\rm Tr}\left[\ h^\dagger T^\alpha h\ +\ k^\dagger T^\alpha k \
\right]
\ \ ,  \nonumber \\ 
&&
{\hat X}_0^\alpha\ =\ {\rm Tr}\left[\ t^\dagger T^\alpha t\ +\ t^\dagger t T^\alpha\ \right]
\ +\ 
2\ {\rm Tr}\left[\ h^\dagger T^\alpha h\ -\ k^\dagger T^\alpha k \
\right]
\ \ .
\end{eqnarray}
One makes the following particle identifications
\begin{eqnarray}
&& 
\rho\ =\ \textstyle{1\over\sqrt{2}}\left[\ h-k\ \right]
\ \ ,\ \ 
a_1\ =\ \cos\psi\ t_1\ +\ \sin\psi 
\textstyle{1\over\sqrt{2}}\left[\ h+k\ \right]\ ,
\nonumber\\
&&
\pi\ =\ -\sin\psi\ t_1\ +\ \cos\psi 
\textstyle{1\over\sqrt{2}}\left[\ h+k\ \right]
\ \ ,\ \ 
f_0 \ =\ t_0
\ \  ,
\end{eqnarray}
which lead to 
\begin{eqnarray}
& &\langle\ \rho^0\ | \ \hat X_0^1 \ -\ i\ \hat X_0^2\ |\ \pi^+\ \rangle \ =\ -\sqrt{2}\cos\psi \ \ ,\ \ 
\langle\ f_0\ | \ \hat X_0^1 \ -\ i\ \hat X_0^2\ |\ \pi^+\ \rangle \ =\
-\sqrt{2}\sin\psi \ , \nonumber \\
& &\langle\ \rho^0\ | \ \hat X_0^1 \ -\ i\ \hat X_0^2\ |\ a_1^+\ \rangle \ =\ -\sqrt{2}\sin\psi \ \ ,\ \ 
\langle\ f_0\ | \ \hat X_0^1 \ -\ i\ \hat X_0^2\ |\ a_1^+\ \rangle \ =\
\sqrt{2}\cos\psi \ , \nonumber \\
&&\qquad\qquad\qquad M_{a1}^2\ =\ M_{\rho}^2\ +\ M_{f_0}^2 \ \ \ , \ \ \  M_{\rho}^2\ =\ \sin^2\psi\ M_{a1}^2
\ \ . 
\end{eqnarray}
The absence of inelastic diffractive scattering, as encapsulated in eq.~(\ref{eq:diff}),
requires that
$\psi={\pi\over 4}$ which, through the decay rate
$\Gamma (\rho\rightarrow\pi\pi )\sim 283\,\cos^2\psi~{\rm MeV}$,
is in excellent agreement with experiment.
This value of the mixing angle is consistent with the large and
uncertain widths of $f_0$ and $a_1$.
(It is interesting that the mixing angle takes a value halfway between
the sigma model scenario ($\psi={\pi\over 2}$) where $\pi$ is paired
with the scalar $f_0$ in a degenerate multiplet, and
the vector-limit scenario~\cite{georgi} ($\psi=0$) where $\pi$ is paired
with the vector $\rho$ in a degenerate multiplet.)
One also obtains the well-known mass relation~\cite{sfsr}
\begin{eqnarray}
&& M_{\rho}^2\ =\ M_{\sigma}^2 \ =\  M_{a1}^2\ -\ M_{\rho}^2 \ .
\end{eqnarray}
The phenomenology of this embedding is discussed in detail in
Refs.~\cite{Fred,Weinberga,Weinbergd,Beane5,Beane7}, and expressed in terms of
chiral perturbation theory parameters in Ref.~\cite{Beane6}.

Next we consider the sector with $G\eta=-1$. The minimal
representation with mass splittings is
$({\bf 2},{\bf 2})\oplus ({\bf 1},{\bf 1})\oplus ({\bf 1},{\bf 1})$
which corresponds to $0^-_-$, $0^-_-$, $0^+_-$ and $1^+_-$ from eq.~(\ref{eq:lightmesonsINTS}).
This chiral representation is composed of the $\lambda=0$
helicity states of $\eta$, $\omega(782)$, $a_0(980)$ and $f_1(1285)$.
In order to construct this chiral multiplet, we introduce the fields
$\lambda_1$, $\lambda_2$ which transform as $({\bf 1},{\bf 1})$
and $y^a_b$ which transforms as a $({\bf 2},{\bf 2})$ under the chiral group,
$y\rightarrow L\ y\ R^\dagger$ with $I=0$ component $y_0$, and  
$I=1$ component, $y_1$.
The free-field dynamics of the helicity-zero states
are determined by the two-dimensional effective Lagrange density
constructed from $\lambda_1$, $\lambda_2$ and $y$,
\begin{eqnarray}
{\cal L}^-_0 & = & 
\partial_+\lambda_1^\dagger \partial_-\lambda^{}_1
\ +\ 
\partial_+\lambda_2^\dagger \partial_-\lambda^{}_2
\ +\ 
{\rm Tr}\left[\ \partial_+ y^\dagger \partial_- y\ \right]\nonumber \\
&&- M_{{\bf 1}y}^2 \ {\rm Tr}\left[\ y^\dagger y\ \right]
\ -\ M_{{\bf 1}\lambda_1}^2 \ \lambda_1^\dagger\lambda^{}_1
\ -\ M_{{\bf 1}\lambda_2}^2 \ \lambda_2^\dagger\lambda^{}_2\nonumber \\
&&\ -\ \left(\ 
{\cal A}_1\ {\rm Tr}\left[\ y^\dagger\ v\ \right]\ \lambda^{}_1\ +\ 
{\cal A}_2\ {\rm Tr}\left[\ y^\dagger\ v\ \right]\ \lambda^{}_2
\ +\ {\rm h.c.}\ \right)
\ \ \ ,
\end{eqnarray}
where ${\cal A}_{1,2}$ are unknown constants,
and the axial-current operator is
\begin{eqnarray}
&&
{\hat X}_0^\alpha\ =\  {\rm Tr}\left[\ y^\dagger T^\alpha
  y\ + \ y^\dagger y T^\alpha\ \right]\ \ .
\end{eqnarray}
Since $\omega$ carries opposite normality to $\eta$ and $f_1$, it
is necessarily unmixed (${\cal A}_{1}=0$) and transforms as a chiral singlet. Hence
we have the particle identifications
\begin{eqnarray}
&& \eta\ =\ \cos\psi\ \lambda_2\ +\ \sin\psi\ y_0 \ \ ,\ \ 
f_1\ =\ -\sin\psi\ \lambda_2\ +\ \cos\psi \ y_0 \ , \nonumber\\
&&\qquad\qquad\quad a_0\ =\ y_1 \ \ ,\ \  \omega\ =\ \lambda_1 \ \ ,
\end{eqnarray}
and we find the following non-trivial solution for the axial coupling constants
and masses,
\begin{eqnarray}
& &\langle\ a_0^+\ | \ \hat X_0^1 \ +\ i\ \hat X_0^2\ |\ f_1\ \rangle \ =\ \sqrt{2}
\cos\psi \ \ ,\ \ 
\langle\ a_0^+\ | \ \hat X_0^1 \ +\ i\ \hat X_0^2\ |\ \eta\ \rangle \ =\ \sqrt{2}
\sin\psi \ , \nonumber \\
&&\quad\qquad\qquad\qquad\qquad  M_{f_1}^2\cos^2\psi\ +\ 
M_{\eta}^2\sin^2\psi\ =\ 
M_{a_0}^2
\ \ \ .
\end{eqnarray}
The masses in this multiplet are well-known experimentally and using the mass
relation to fix the mixing angle gives $\psi=44.5^o$,
in excellent agreement with the inelastic
diffraction constraint imposed by eq.~(\ref{eq:diff}). The predicted value for the $f_1$ width,
$\Gamma (f_1\rightarrow a_0\pi )\sim 76\,\cos^2\psi~{\rm MeV}$, is then
quite respectable. The value for the $a_0$ width,
$\Gamma (a_0\rightarrow\eta\pi )\sim 647\,\sin^2\psi~{\rm MeV}$, would appear
to over predict the (rather uncertain) experimental width. However, as in the
case of the $a_1$ and $f_0$, one should not expect to do better in the scalar
sector given the assumptions that have been made.

Finally, we consider the remaining degrees of freedom necessary to fill out
the interpolators of eq.~(\ref{eq:lightmesonsINTS}): the
$\lambda= 1$ ($\uparrow$) components of $\rho$, $a_1$,
$\omega$ and $f_1$. Evidently, these states are in two decoupled 
$({\bf 2},{\bf 2})$ representations. One immediately finds
\begin{eqnarray}
& &\langle\ \rho^+\ | \ \hat X_\uparrow^1 \ +\ i\ \hat X_\uparrow^2\ |\ \omega \ \rangle \ =\ \sqrt{2}     \ \ ,\ \ 
\langle\ a_1^+\ | \ \hat X_\uparrow^1 \ +\ i\ \hat X_\uparrow^2\ |\ f_1 \ \rangle \ =\ \sqrt{2} \ ,\nonumber \\
&&\qquad\qquad\qquad\qquad  M_\rho^2\ =\ M_\omega^2  \ \ ,\ \  M_{a_1}^2\ =\ M_{f_1}^2 
\ \ \ ,
\end{eqnarray}
and hence the masses of the $G\eta=\pm 1$ sectors are related by eq.~(\ref{eq:comma}).
While the mass relations are in excellent agreement with data, the axial
couplings cannot be confronted with data at present.

\section{The Heavy Mesons}
\label{sec:heavyM}

Heavy mesons fall into degenerate doublets labeled by
$j_\pm^P=(j_\ell\pm{1\over 2})^{P_\ell}$, where $P$ ($P_\ell$) is the parity of
the heavy meson (light degrees of freedom) and $j$ ($j_\ell$) is the angular
momentum of
the heavy meson (light degrees of freedom). The ground-state mesons have
$j_\ell={1\over 2}$ and $P_\ell=(-)$ and are denoted $P$ ($0^-$) and $P^*$
($1^-$), while the first excited heavy-meson doublet also have $j_\ell={1\over 2}$
and $P_\ell=(+)$, and are denoted $P_0^*$ ($0^+$) and $P_1^\prime$ ($1^+$).
Consider first the $\lambda=0$ heavy mesons.  The axial couplings between the
$\lambda=0$ states are defined as~\cite{Falk}
\begin{eqnarray}
&&\qquad\qquad\langle\ P  \ | \ \hat X_0^\alpha \ |\ P^*\ \rangle \ =\ g\ T^\alpha \ ,\nonumber \\
&&\langle\ P  \ | \ \hat X_0^\alpha \ |\ P_0^*\ \rangle \ =\ 
\langle\ P_1^\prime  \ | \ \hat X_0^\alpha \ |\ P^*\ \rangle \ =\ h\ T^\alpha \ \ ,
\nonumber \\
&& \qquad\qquad \langle\ P_1^\prime  \ | \ \hat X_0^\alpha \ |\ P_0^*\ \rangle \ =\ g^\prime\ T^\alpha
\ \ .
\end{eqnarray}
Generally the heavy mesons are sums of any number of $({\bf 2},{\bf 1})$ and
$({\bf 1},{\bf 2})$ representations.  This case is particularly easy to
analyze. Weinberg has shown that $|g|\ \leq\ 1$~\cite{Weincq}, a bound which
holds for all axial transitions among all heavy mesons~\cite{Beane1}. If one imposes the
constraint of no inelastic diffraction, then Weinberg has further shown that
$|g|=0,1$ are the only possibilities. 
Heavy meson interpolators with space-time quantum numbers, $J^{P}$, of
pseudoscalar, vector, axialvector and scalar character decompose to
\begin{eqnarray}
(\ \overline{Q}^{\, a} q_a\ )_{P}\ &\rightarrow &\ \textstyle{1\over 2}^- \ \ , \ \ 
(\ \overline{Q}^{\, a} q_a\ )_{V}\ \rightarrow\  \textstyle{1\over 2}^+ \ , \nonumber \\
(\ \overline{Q}^{\, a} q_a\ )_{S}\ &\rightarrow &\ \textstyle{1\over 2}^+ \ \ , \ \ 
(\ \overline{Q}^{\, a} q_a\ )_{A}\ \, \rightarrow\ \textstyle{1\over 2}^- \ ,
\label{eq:heavymesonsINTS}
\end{eqnarray}
where we have used the schema $I^\eta$. These quantum numbers
correspond to $P$, $P^*$, $P_0^*$ and $P_1^\prime$, respectively.
Now in order to construct
states of definite normality we require at least the representation
$({\bf 2},{\bf 1})\oplus ({\bf 1},{\bf 2})$,
and we might naively place the
ground state heavy mesons in this representation. 
HQSS then requires that $M_{\bf 22}^2$ vanish between these states 
and the absence of the $P\rightarrow P$ (required by normality)
matrix element yields $|g|=1$, which is
the NCQM value. Placing the excited doublet in an analogous multiplet then gives
$|g^\prime|=1$ and $|h|=0$ as well. However, there is a subtlety with these 
embeddings. As one moves away from the
heavy-quark limit and turns on $M_{\bf 22}^2$, one finds that the $P$ and $P^*$
masses are of the form $M_{\bf 1}^2\ \pm\  M_{\bf 22}^2$. 
However, 
the form of the leading spin-symmetry violating operator, $\sigma\cdot G$ with
matrix elements parameterized by $\lambda_2$,
requires
that the spin-averaged combination $3\ M_{P^*}^2\ +\ M_{P}^2$ be {\it independent} of 
$M_{P^*}^2\ -\ M_{P}^2$, and  there is an analogous constraint for the 
excited doublet. Hence one expects that the ground- and first excited-state
heavy mesons fill out a reducible
$({\bf 2},{\bf 1})\oplus ({\bf 1},{\bf 2})\oplus({\bf 2},{\bf 1}) \oplus ({\bf 1},{\bf 2})$ 
representation. It is straightforward to show that
\begin{eqnarray}
&& g\ =\ \cos ( \theta +\phi )\ =\ -g^\prime 
\ \ ,\ \ 
h\ = \  \sin ( \theta +\phi ) \ \ ,
\label{eq:heavymesonsAXIALS}
\end{eqnarray}
where $\theta$ ($\phi$) is the mixing angle between the states with $\eta=-1$ ($\eta=+1$).
The constraint of no inelastic diffraction implies $\theta=\phi={\pi\over 4}$
and one finds $|g|=0$, $|h|=1$, consistent with the theorem of Ref.~\cite{Weincq}.

Finally, we have the remaining degrees of freedom necessary to fill out
the interpolators of eq.~(\ref{eq:lightmesonsINTS}): the
$\lambda= 1$ ($\uparrow$) components of $P^*$ and $P_1^\prime$. The axial
matrix elements between these states are
\begin{eqnarray}
&&\langle\ P^*  \ | \ \hat X_\uparrow^\alpha \ |\ P^*\ \rangle \ =\ g \ T^\alpha \ \ ,\ \
\langle\ P_1^\prime  \ | \ \hat X_\uparrow^\alpha \ |\ P_1^\prime\ \rangle \ =\
g^\prime \ T^\alpha \ \ , \nonumber \\
&& \qquad\qquad \langle\ P_1^\prime  \ | \ \hat X_\uparrow^\alpha \ |\ P^*\ \rangle \ =\ h\ T^\alpha
\ \ .
\end{eqnarray}
Consistency with eq.(\ref{eq:comma}) and 
eq.~(\ref{eq:heavymesonsAXIALS}) requires that the 
$\lambda= 1$ components of $P^*$ and $P_1^\prime$ be in a reducible
$({\bf 2},{\bf 1})\oplus ({\bf 1},{\bf 2})$ with a mixing angle given
by $(\theta +\phi )/2$ which takes the value ${\pi\over 4}$ when the
inelastic diffraction constraint of eq.~(\ref{eq:diff}) is imposed.

Experimentally, the coupling $g=g_\pi$ is known to be smaller than unity
through different determinations.
The width of the $D^*$, combined with the branching fraction for $D^*\rightarrow D\pi$,
gives $g=0.59\pm 0.01\pm 0.07$~\cite{Ahmed:2001xc} at tree-level in the 
chiral expansion, which is consistent 
with recent lattice QCD calculations~\cite{Yamada:2002wh},
and with a determination from 
$D^*\rightarrow D\pi$ and $D^*\rightarrow D\gamma$
at order $\sqrt{m_q}$ in $SU(3)$ chiral perturbation theory~\cite{Amundson:1992yp}.
At order $m_q$ and $1/m_c$ in chiral perturbation theory~\cite{Stewart:1998ke}
there are two solutions for $g$ derived from $D^*\rightarrow D\pi$
and $D^*\rightarrow D\gamma$.
One solution, 
$g=0.76\pm 0.03\pm 0.15$~\cite{Stewart:1998ke},
is significantly larger than our prediction and it is unlikely
that the discrepancy can be accounted for through chiral corrections.
However, the second solution, $g=0.27\pm 0.03\pm 0.04$,
is consistent with our conjecture and is conceivably
non-zero because of quark mass effects alone.
A small value of $g$ has also been found by Bardeen and Hill who developed
a dynamical model that incorporates a chiral multiplet structure
very similar to that found here~\cite{Bardeen}.

\section{Conclusions}

The observed high-energy behavior of pion-hadron scattering amplitudes, together with the
assumption that pole graphs dominate the amplitude at low energies, is sufficient to provide
algebraic constraints on the structure of hadronic axial-current matrix elements and
mass matrices~\cite{Fred,Weinberga,Weinbergb,Weinbergc,Weinbergd}. These
constraints allow one to prove that the $\pi$, $\rho$, $f_0$ and $a_1$
form a decoupled chiral multiplet in the large-$\nc$ limit of
QCD~\cite{Weinbergd}. We have conjectured that the
entire hadron spectrum is composed of decoupled reducible chiral representations
and we have given a prescription for finding these multiplets. 
Our conjecture provides a map of the chiral structure of the QCD ground state.
In summary, we find:
\begin{eqnarray}
qqq:\qquad  &&\left\{ \begin{array}{r@{\qquad  \qquad}l}
\lambda =\textstyle{1\over 2}\qquad :  &  ({\bf 2},{\bf 3})\oplus ({\bf  1},{\bf 2}) \\
 &  \nonumber \\
\lambda =\textstyle{3\over 2}\qquad :  &  ({\bf 1},{\bf 4}) \end{array} \right. 
\nonumber \\ 
 &  \nonumber \\
Qqq:\qquad  &&\left\{ \begin{array}{r@{\qquad  \qquad}l}
\lambda = \textstyle{1\over 2}\qquad : &  ({\bf 2},{\bf 2})\oplus ({\bf 1},{\bf 3})\oplus ({\bf 1},{\bf 1} ) \\
 & \nonumber  \\
\lambda = \textstyle{3\over 2}\qquad : &  ({\bf 1},{\bf 3}) \end{array} \right. 
\nonumber\\ 
 &  \nonumber \\
\overline{q}q:\qquad  &&\left\{ \begin{array}{r@{\qquad  \qquad}l}
\lambda =0 \qquad : &  ({\bf 2},{\bf 2})\oplus ({\bf 1},{\bf 3})\oplus ({\bf
  3},{\bf 1})\ \ (G\eta=+1) \nonumber\\ 
& ({\bf 2},{\bf 2})\oplus ({\bf 1},{\bf 1})\oplus ({\bf 1},{\bf 1} )\ \ (G\eta=-1) \\
\lambda =1\qquad :  & ({\bf 2},{\bf 2})\oplus ({\bf 2},{\bf 2}) \end{array} \right.  
\nonumber\\ 
 & \nonumber  \\
\overline{Q}q:\qquad  &&\left\{ \begin{array}{r@{\qquad  \qquad}l}
\lambda =0\qquad :  & ({\bf 1},{\bf 2})\oplus ({\bf 2},{\bf 1})\oplus ({\bf 1},{\bf 2})\oplus ({\bf 2},{\bf 1}) \\
 &   \\
\lambda =1\qquad :  & ({\bf 1},{\bf 2})\oplus ({\bf 2},{\bf 1}) \end{array}
\right. 
\nonumber
\end{eqnarray}
for ground-state light baryons, heavy baryons, light mesons and heavy mesons,
respectively.  Evidently, the chiral multiplet structure that nature has chosen
contains the minimal particle content necessary to saturate the interpolating
fields for the hadrons, and to allow for non-zero mass splittings between
members of the multiplet.  The claim that chiral multiplets are small and
decoupled might appear odd given that there are observed axial transitions
---say in the light-baryon sector--- from excited states to the ground-state
multiplet.  However, the decoupled chiral multiplets mix when the quark masses
are turned on. Therefore, the smallness of the axial transitions from excited
multiplets to the ground-state multiplet as compared to those within the
ground-state multiplet is due to the smallness of the quark masses.

Our conjecture agrees with all experimental data available for the low-lying
hadrons and leads to new and exciting predictions.  We look forward to
exploration of this conjecture both theoretically and experimentally.

\acknowledgements
We thank David Kaplan, Aneesh Manohar, Roxanne Springer and Iain Stewart for valuable discussions.
This work is supported in part by the U.S. Department of Energy under 
Grant No.~DE-FG03-00-ER-41132 (SRB) and Grant No.~DE-FG03-97-ER-4014 (MJS).

\end{document}